\documentclass[twoside,onecolumn,11pt]{article}

\usepackage{blindtext} 

\usepackage[sc]{mathpazo} 
\usepackage[T1]{fontenc} 
\usepackage{microtype} 
\usepackage[english]{babel} 
\usepackage[hmarginratio=1:1,top=32mm,columnsep=20pt]{geometry} 
\usepackage[hang, small,labelfont=bf,up,textfont=it,up]{caption} 
\usepackage{booktabs} 
\usepackage{lettrine} 
\usepackage{enumitem} 
\usepackage{lineno,hyperref,color}
\usepackage{epstopdf}
\usepackage{soul}
\usepackage{amsmath}
\usepackage{booktabs}
\usepackage{tikz}
\usepackage{pgf}
\usepackage{xcolor}
\usetikzlibrary{arrows,shapes,snakes,automata,backgrounds,petri}
\usetikzlibrary{positioning}
\usepackage{graphicx}
\usepackage{sidecap}
\usepackage{multicol}
\usepackage{lipsum}
\usepackage{stfloats}
\setlist[itemize]{noitemsep} 
\usepackage{abstract} 

\usepackage{titlesec} 
\renewcommand\thesection{\Roman{section}} 
\renewcommand\thesubsection{\roman{subsection}} 
\titleformat{\section}[block]{\large\scshape\centering}{\thesection.}{1em}{} 
\titleformat{\subsection}[block]{\large}{\thesubsection.}{1em}{} 

\usepackage{fancyhdr} 
\usepackage{titling} 

\usepackage{hyperref} 

\tikzset{
  block/.style    = {draw, thick, rectangle, minimum height = 2em,minimum width = 3em},
  sum/.style      = {draw, circle},
  comp/.style      = {draw, circle},
  text/.style    = {draw},
  line/.style   = {draw},
  input/.style   = {draw},
  output/.style   = {draw},
  decision/.style= {diamond, minimum width=3cm, minimum height=1cm, text centered, draw=black, fill=white!30, aspect=3}
}

\modulolinenumbers[5]

\newcommand{\todo}[1]{}
\renewcommand{\todo}[1]{{\color{red} TODO: {#1}}}
 \newcommand{\virg}[1]{``#1''}

\pretitle{\begin{center}\Huge\bfseries} 
\posttitle{\end{center}} 
\title{A Hybrid Approach to\\ Video Source Identification} 
\author{%
\textsc{Massimo Iuliani, Marco Fontani, Dasara Shullani, and~Alessandro Piva}\thanks{M. Iuliani, M. Fontani and A. Piva are also with FORLAB, Multimedia
Forensics Laboratory, PIN Scrl, 59100 Prato, Italy} \\[1ex] 
\normalsize Department of Information Engineering, University of
Florence, 50139 Florence, Italy \\ 
\normalsize \href{mailto:massimo.iuliani@unifi.it}{massimo.iuliani@unifi.it} 
}
\renewcommand{\maketitlehookd}{%
\begin{abstract}
\noindent 
Multimedia Forensics allows to determine whether videos or images
have been captured with the same device, and thus, eventually, by
the same person. Currently, the most promising technology to
achieve this task, exploits the unique traces left by the camera
sensor into the visual content. Anyway, image and
video source identification are still treated separately from one
another. This approach is limited and anachronistic if we consider
that most of the visual media are today acquired using
smartphones, that capture both images and videos. In this paper we
overcome this limitation by exploring a new approach that allows
to synergistically exploit images and videos to study the device
from which they both come. Indeed, we prove it is possible to
identify the source of a digital video by exploiting a reference
sensor pattern noise generated from still images taken by the same
device of the query video. The proposed method provides comparable
or even better performance, when compared to the current video
identification strategies, where a reference pattern is estimated
from video frames. We also show how this strategy can be effective
even in case of in-camera digitally stabilized videos, where a
non-stabilized reference is not available, by solving some state-of-the-art limitations. We explore a possible direct application of this result, that is social media profile linking, i.e. discovering relationships between two or more social media profiles by comparing the visual contents - images or videos - shared therein.

\end{abstract}
}

\begin{document}

\maketitle

\section{Introduction}
Digital videos (DVs) are steadily becoming the preferred means for
people to share information in an immediate and convincing way.
Recent statistics showed a 75\% increase in the number of DVs
posted on Facebook in one year~\cite{facebook} and posts
containing DVs yields more engagement than their text-only
counterpart~\cite{facebook2}. Interestingly, the vast majority of
such contents are captured using smartphones, whose impact on
digital photography is dramatic: in 2014, compact camera sales
dropped by 40\% worldwide, mostly because they are being replaced
by smartphone cameras, which are always at your fingertips and
makes sharing much easier~\cite{smartvscamera}.

In such a scenario, it is not surprising that digital videos
gained importance also from the forensic and intelligence point of
view: videos have been recently used to spread terror over the
web, and many critical events have been filmed and shared by
thousands of web users. In such cases, investigating the digital
history of DVs is of paramount importance in order to recover
relevant information, such as acquisition time and place,
authenticity, or information about the source device. In the last
decades Multimedia Forensics has developed tools for such tasks,
based on the observation that each processing step leaves a
distinctive trace on the digital content, as a sort of digital
fingerprint. By detecting the presence, the absence or the
incongruence of such traces it is possible to blindly investigate
the digital history of the content~\cite{piva2013overview}.

In particular, the source identification problem - that is,
univocally linking the digital content to the device that captured
it - received great attention in the last years. Currently, the
most promising technology to achieve this task exploits the
detection of the sensor pattern noise (SPN) left by the
acquisition device~\cite{lukas2006digital}. This footprint is
universal (every sensor introduces one) and unique (two SPNs are
uncorrelated even in case of sensors coming from two cameras of
same brand and model). As long as still images are concerned, SPN
has been proven to be robust to common processing operations like
JPEG compression~\cite{lukas2006digital}, or even uploading to
social media platforms
(SMPs)~\cite{castiglione2013experimentations, bertini2016social}.

On the contrary, research on source device identification for DVs
is not as advanced. This is probably due to the higher
computational and storage effort required for video analysis, the
use of different video coding standards, and the absence of
sizeable datasets available to the community for testing. Indeed,
DV source identification borrowed both the mathematical background
and the methodology from the still image
case~\cite{chen2007source}: like for images, thus, assessing the
origin of a DV requires the analyst to have either the source
device or some training DVs captured by that device, from which to
extract the reference SPN.

However, if we consider that $85\%$ of shared media are captured
using smartphones, which use the same sensor to capture both images
and videos, it is possible to exploit images also for video source
identification. A first hint in using still images to estimate the
video fingerprint was recently provided
in~\cite{2017stabilisedvideo}, where the authors noticed how image
and video patterns of some portable devices acquiring
non-stabilized video can be generally related by cropping and
scaling operations. Anyway, in the research community, there's
still no better way to perform image and video source
identification than computing two different reference SPNs, one
for still images and one for videos respectively. In addition, a
strong limitation is represented by the presence in many mobile
devices of an in-camera digital video stabilization algorithm,
such that a non-stabilized SPN reference cannot be estimated from
a DV~\cite{chen2007source}.

The first contribution of this work focuses on proposing a hybrid
source identification approach, that exploits still images for estimating the fingerprint that will be used to verify the source of a video. The geometrical relation between image and video
acquisition processes are studied for $18$ modern smartphones,
including devices with in-camera digital stabilization. Secondly,
we prove that the proposed technique, while preserving the state of the
art performance for non-stabilized videos, is able to effectively
detect the source of in-camera digitally stabilized videos also.
Furthermore, this hybrid approach is used to link image and video
contents belonging to different social media platforms,
specifically Facebook and YouTube. 

The rest of the paper is organized as follows: Section \ref{sec:soa} introduces SPN based source device identification, and reviews the state of the art for DV source identification; Section \ref{sec:hybrid} formalizes the considered problem and describes the proposed hybrid approach; 
Section \ref{sec:dataset} presents the video dataset prepared for the tests, and discusses some YouTube/Facebook technical details related to the SPN; Section \ref{sec:experiments} is dedicated to the experimental validation of the proposed technique, including comparison with existing approaches, and tests on stabilized videos and on contents belonging to SMPs; finally, Section \ref{sec:conclusion} draws some final remarks and outlines future works.

Everywhere in this paper vectors and matrices are denoted in bold as $\mathbf{X}$ and their components as $\mathbf{X}(i)$ and $\mathbf{X}(i,j)$ respectively. All operations are element-wise, unless mentioned otherwise. Given two vectors $\mathbf{X}$ and $\mathbf{Y}$ we denote as $||\mathbf{X}||$ the euclidean norm of $\mathbf{X}$, as $\mathbf{X \cdot Y}$ the dot product between $\mathbf{X}$ and $\mathbf{Y}$, as $\bar{\mathbf{X}}$ the mean values of $\mathbf{X}$, as $\rho(s_1,s_2; \mathbf{X},\mathbf{Y})$ the normalized cross-correlation between $\mathbf{X}$ and $\mathbf{Y}$ calculated as

{ \small \begin{eqnarray*} \rho(s_1,s_2; \mathbf{X},\mathbf{Y}) = \frac{\sum_{i} \sum_{j} ( \mathbf{X}(i,j) - \bar{\mathbf{X}} ) ({\mathbf{Y}(i+s_1,j+s_2) - \bar{\mathbf{Y}}} )} {||\mathbf{X} - \bar{\mathbf{X}}|| ||\mathbf{Y} - \bar{\mathbf{Y}}||}  \end{eqnarray*} } \\
If $\mathbf{X}$ and $\mathbf{Y}$ dimensions mismatch, a zero down-right padding is applied.
Furthermore its maximum, namely the $\max\limits_{\mathbf{s_1,s_2}} \rho(s_1,s_2; \mathbf{X},\mathbf{Y})$, is denoted as $\rho_{peak}(\mathbf{X},\mathbf{Y}) = \rho(\mathbf{s}_{peak}; \mathbf{X},\mathbf{Y})$.
The notations are simplified in $\rho(s_1,s_2)$ and in $\rho_{peak}$ when the two vectors cannot be misinterpreted.

\section{Digital Video Source Device Identification Based on Sensor Pattern Noise}
\label{sec:soa}
The task of blind source device identification has gathered great attention in the multimedia forensics community. Several approaches were proposed to characterize the capturing device by analyzing traces like sensor dust \cite{dirik2008digital}, defective pixels \cite{geradts2001methods}, color filter array interpolation \cite{bayram2005source}. A significant breakthrough was achieved when Lukas et al. first introduced the idea of using Photo-Response Non-Uniformity (PRNU) noise to univocally characterize camera sensor \cite{lukas2006digital}. Being a multiplicative noise, PRNU cannot be effectively removed even by high-end devices; moreover, it remains in the image even after JPEG compression at average quality. 
The suitability of PRNU-based camera forensics for images
retrieved from common SMPs has been investigated in
\cite{castiglione2013experimentations}, showing that modifications
applied either by the user or by the SMP can make the source
identification based on PRNU ineffective. The problem of
scalability of SPN-based camera identification has been
investigated in several works \cite{goljan2009large,
cattaneo2014scalable}. Noticeably, in \cite{goljan2009large}
authors showed that the Peak-to-Correlation Energy (PCE) provides
a significantly more robust feature compared to normalized
correlation. The vast interest in this research field fostered the
creation of reference image databases specifically tailored for
the evaluation of source identification \cite{gloe2010dresden},
allowing a thorough comparison of different methods
\cite{liu2015enhancing}. Recently, authors of
\cite{valsesia2015compressed} addressed the problem of reducing
the computational complexity of fingerprint matching, both in
terms of time and memory, through the use of random projections to
compress the fingerprints, at the price of a small reduction in
matching accuracy.

All the methods mentioned so far have been thought for (and tested
on) still images. Although research on video source identification
began almost at the same time, the state of the art is much
poorer. In their pioneering work \cite{chen2007source}, Chen et
al. proposed to extract the SPN from each frame separately and
then merge the information through a Maximum Likelihood Estimator;
as to the fingerprint matching phase, the PCE was recommended
\cite{chen2007source}. The experimental results showed that
resolution and compression have an impact on performance, but
identification is still possible if the number of considered
frames can be increased (10 minutes for low resolution, strongly
compressed videos). Two years later, Van Houten et al.
investigated the feasibility of camcorder identification with
videos downloaded from YouTube \cite{van2009source}, yielding
encouraging results: even after YouTube recompression, source identification was
possible. However, results in \cite{van2009source} are outdated,
since both acquisition devices and video coding algorithms have
evolved significantly since then. This study was extended by
Scheelen et al. \cite{scheelen2012camera}, considering more recent
cameras and codecs. Results confirmed that source identification is
possible, however authors clarify that the reference pattern was
extracted from reference and natural videos before re-encoding.
Concerning reference pattern estimation, Chuang et al.
\cite{chuang2011exploring} firstly proposed to treat differently
the SPN extracted from video frames based on the type of their
encoding; the suggested strategy is to weigh differently intra-
and inter- coded frames, based on the observation that intra-coded
frames are more reliable for PRNU fingerprint estimation, due to
less aggressive compression. A recent contribution from Chen et
al. \cite{chen2015live} considered video surveillance systems, where the videos 
transmitted over an unreliable wireless channel, can be affected by blocking artifacts, complicating
pattern estimation.

Most of the research on video forensics neglects the analysis of
digitally stabilized videos, where the SPN can be hardly
registered. In~\cite{hoglund2011identifying} an algorithm was
proposed to compensate the stabilization on interlaced videos.
Anyway, the method was tested on a single device and it is inapplicable on the vast majority of modern devices, that come to life with a $1080p$ camera ($p$ stands for \textit{progressive}).
Recently, Taspinar et al. ~\cite{2017stabilisedvideo} showed that digital
stabilization applied out of camera by a third party program can
be managed by registering all video frame on the first frame based
on rotation and scaling transformation. Anyway the technique is
proved to be really effective only when a reference generated
from non-stabilized videos is available.
This is a gap to be filled considering that most modern smartphones features in-camera digital stabilization, and that in many cases such feature cannot be disabled.\\

As the reader may have noticed, all the mentioned works discuss
source identification \textit{either} for still images or videos (with the only exception of ~\cite{2017stabilisedvideo}),
and in the vast majority of cases the reference pattern is
estimated from \virg{clean} contents, meaning images or frames as
they exit from the device, without any alteration due to
re-encoding or (even worse) upload/download from SMPs. This
approach seriously limits the applicability of source device
identification, since it assumes that either the device or some
original content is available to the analyst.
In the following sections we show how to exploit the available
mathematical frameworks to determine the source of a DV based on
a reference derived by still images, even in the case of in-camera
digitally stabilized videos, and eventually apply this strategy to
link images and video from different SMPs.

\section{Hybrid Sensor Pattern Noise Analysis}
\label{sec:hybrid}

\begin{figure}
    \begin{center}
\includegraphics[width=0.8\columnwidth]{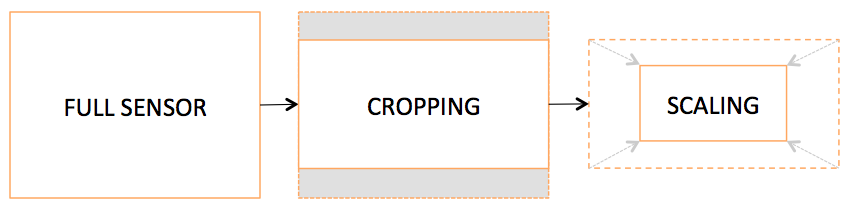}
    \caption{Example of geometric transformation in video acquisition.}
    \label{HybridDevices}
    \end{center}
\end{figure}

Digital videos are commonly captured at a much lower resolution
than images: top-level portable devices reach 4K video resolution
at most (which means, 8 Megapixels per frame), while the same
devices easily capture 20 Megapixels images. During video
recording, a central crop is carried so to adapt the sensor size
to the desired aspect ratio (commonly 16:9 for videos), then the
resulting pixels are scaled so to match exactly the desired
resolution (see Figure \ref{HybridDevices}). As a direct
consequence, the sensor pattern noise extracted from images and
videos cannot be directly compared and most of the times, because
of cropping, it is not sufficient to just scale them to the same
resolution.
The hybrid source identification (HSI) process consists in
identifying the source of a DV based on a reference derived from
still images. The strategy involves two main steps: i) the
reference fingerprint is derived from still images acquired by the
source device; ii) the query fingerprint is estimated from the
investigated video and then compared with the reference to verify
the possible match.

The camera fingerprint $\mathbf{K}$ can be estimated from $N$
images $\mathbf{I}^{(1)}, \dots,  \mathbf{I}^{(N)}$ captured by
the source device. A denoising filter \cite{lukas2006digital},
\cite{mihcak1999spatially} is applied to each frame and the noise
residuals $\mathbf{W}^{(1)}, \dots,  \mathbf{W}^{(N)}$ are
obtained as the difference between each frame and its denoised
version. Then the fingerprint estimation $\widetilde{\mathbf{K}}$
is derived by the maximum likelihood
estimator~\cite{chen2008determining}:
\begin{equation}
\widetilde{\mathbf{K}} = \frac{\sum_{i=1}^N \mathbf{W}^{(i)} \mathbf{I}^{(i)}}{\sum_{i=1}^N (\mathbf{I}^{(i)})^2}.
\label{eq:lr}
\end{equation}
The fingerprint of the video query is estimated in the same way by the available video frames. \\
Denoting by $\widetilde{\mathbf{K}}_R$ and
$\widetilde{\mathbf{K}}_Q$ the reference and query fingerprints,
the source identification is formulated as a two-channel
hypothesis testing problem~\cite{holt1987two}
\begin{align*}
H_0 : \mathbf{K}_R \neq \mathbf{K}_Q \\
H_1 : \mathbf{K}_R = \mathbf{K}_Q.
\end{align*}
where $\widetilde{\mathbf{K}}_R = \mathbf{K}_R + \Xi_R$ and
$\widetilde{\mathbf{K}}_Q = \mathbf{K}_Q + \Xi_Q$, being $\Xi_R$
and $\Xi_Q$ noise terms. In the considered case,
$\widetilde{\mathbf{K}}_R$ and $\widetilde{\mathbf{K}}_Q$ are
derived from still images and video frames respectively, thus
differing in resolution and aspect ratio due to the cropping and
resize occurring during the acquisition (see
Fig. \ref{HybridDevices}). Then, the test statistic is built as
proposed in ~\cite{goljan2008camera}, where the problem of camera
identification from images that were simultaneously cropped and
resized was studied: the two-dimensional normalized
cross-correlation $\rho(s_1,s_2)$ is calculated for each of the
possible spatial shifts $(s_1,s_2)$ determined within a set of
feasible cropping parameters. Then, given the peak $\rho_{peak}$,
its sharpness is measured by the Peak to Correlation Energy (PCE)
ratio~\cite{goljan2009large} as
\begin{equation}
\label{eq:PCE}
PCE = \frac{\rho(\mathbf{s}_{peak})} { \frac{1}{mn - |\mathcal{N}|} \sum\limits_{\mathbf{s}\notin\mathcal{N}} \rho(\mathbf{s}) }
\end{equation}
where $\mathcal{N}$ is a small set of peak neighbors. \\
In order to consider the possible different scaling factors of the
two fingerprints - since videos are usually resized - a brute
force search can be conducted considering the PCE as a function of
the plausible scaling factors $r_0, \dots, r_m$. Then its maximum
\begin{equation}
\label{eq:p}
P = \max\limits_{r_i} PCE(r_i)
\end{equation}
is used to determine whether the two fingerprints belong to the
same device. Practically, if this maximum overcomes a threshold
$\tau$, $H_1$ is decided and the corresponding values
$\mathbf{s}_{peak}$ and $r_{peak}$ are exploited to determine the
cropping and the scaling factors. In ~\cite{goljan2008camera} it
is shown that a theoretical upper bound for False Alarm Rate (FAR)
can be obtained as
\begin{equation}
\label{eq:FAR}
FAR = 1-(1-Q(\sqrt{\tau}))^k
\end{equation}
where $Q$ is the cumulative distribution function of a normal variable N(0,1) and $k$ is the number of tested scaling and cropping parameters. \\
This method is expected to be computationally expensive, namely for large dimension images. Anyway, this problem can be mitigated considering that:
\begin{itemize}
\item if the source device is available, or its model is known, the resize and cropping factors are likely to be determined by the camera software specifics or by experimental testing;
\item even when no information about the model is available, it is not necessary to repeat the whole search on all frames. Once a sufficiently high correlation is found for a given scale, the search can be restricted around it.
\end{itemize}
In Section \ref{sec:dataset}, cropping and scaling factors for $18$ devices are reported. \\

\subsection{Source Identification of Digitally Stabilized Videos}
\label{sec:digtalstab}
    Recent camera softwares include digital stabilization technology to reduce the impact of shaky hands on captured videos. By estimating the impact of the user movement the software adjusts which pixels on the camcorder's image sensor are being used. Image stabilization can be usually turned on and off by the user on devices based on Android OS while in iOS devices this option cannot be modified by the camera software.
The source identification of videos captured with active digital
stabilization cannot be accomplished based on the PRNU
fingerprint: in fact the process disturbs the fingerprints
alignment that is a \emph{sine qua non} condition for the
identification process. HSI solves the problem on the reference
side (the fingerprint is estimated from still images) but the
issue remains on the query side. A first way to compensate digital
stabilization was proposed in~\cite{2011imagestabilisation} and
tested on a single Sony device. Recently,
in~\cite{2017stabilisedvideo}, it was proposed to compute the
fingerprint from a stabilized video by using the first frame noise
as reference and by registering all following frame noises on the
first one by estimating the similarity transformation that
maximize the correlation between the two patterns. The technique
was proved to compensate digital stabilization applied out of
camera by third party software, but with limited reliability. HSI allows to
intuitively perform source identification of stabilized videos: on the reference side, still images are exploited to
estimate a reliable, stable fingerprint, while on the query side, each video
frame is registered on the image reference based on a similarity
transformation. In Section~\ref{subs:res_stab} we will prove the
effectiveness of this technique by estimating the fingerprint with
only five frames on in-camera stabilized videos from modern
devices. In the next section we define the hybrid source
identification pipeline conceived to reduce false alarm and
computational effort.

\subsection{HSI Pipeline}
\begin{figure}
    \begin{center}
\includegraphics[width=\columnwidth]{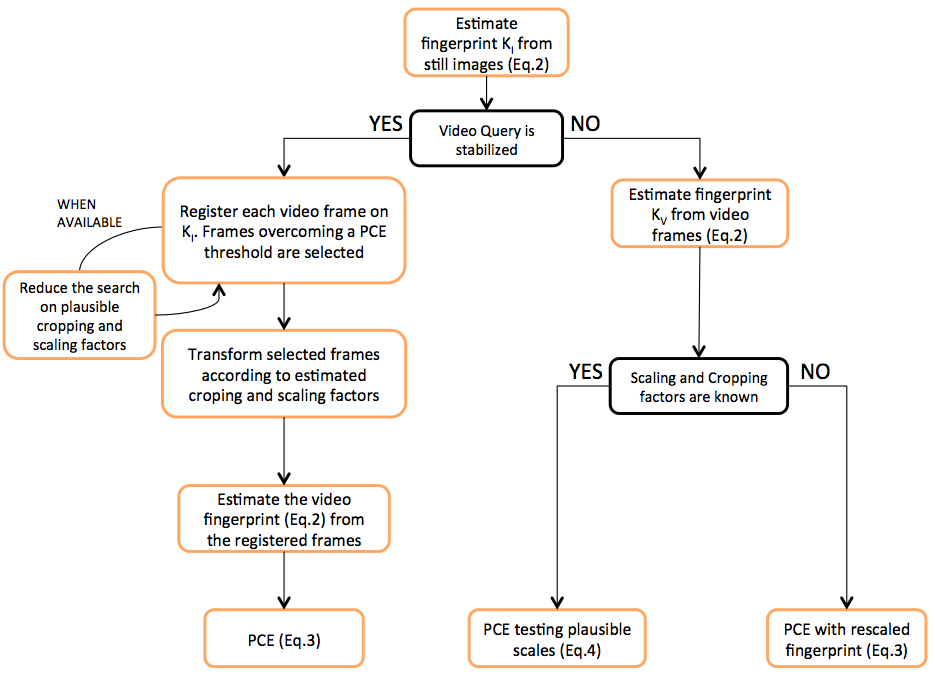}
    \caption{HSI pipeline to source attribution of a query video.}
    \label{HSIpipeline}
    \end{center}
\end{figure}

Given a query video and a set of images belonging to a reference
device, the proposed pipeline is summarized in
Fig.~\ref{HSIpipeline}.
First, the device fingerprint $\mathbf{K}_{I}$ is estimated from
still images according to Eq.~(\ref{eq:lr}). Then, stabilized
videos are preliminary identified by splitting the frames in two
groups that are used independently to estimate two different
fingerprints, as described in \cite{2017stabilisedvideo}, and
computing their PCE; a low PCE value will expose the presence of
digital stabilization. If no stabilization is detected, the video
fingerprint $\mathbf{K}_{V}$ is just estimated treating video frames as still images.
Conversely, each frame is registered on the reference
$\mathbf{K}_{I}$ searching the plausible parameters based on PCE
values. In case the expected range of parameters are known, the
search can be reduced to save computational effort and mitigate
the false alarm (see Section~\ref{sec:fingmatch} for details). Only
the registered video frames overcoming a PCE threshold $\tau$ are then
aggregated to estimate the video fingerprint $\mathbf{K}_{V}$.
Once both fingerprints $\mathbf{K}_{I}$ and $\mathbf{K}_{V}$ are
available, they are compared according to Eq.~(\ref{eq:p}) by testing
plausible scaling factors.
Again, the analysis can be reduced to expected cropping and scaling factors. \\
\subsection{Extension to contents shared on social media platforms}
\label{sec:ProfileLinking}

The proposed technique can be applied to match multimedia contents exchanged through different SMPs.
Let us consider a user publishing, with an anonymous profile, videos with criminal content through a SMP. At the same time this user, say Bob, is leading his virtual social life on another social network where he publicly shares his everyday's pictures.
Unaware of the traces left by the sensor, he captures with the same device the contents he shares on both profiles. Then, the fingerprints derived from the images and videos on the two social platforms can be compared with the proposed method to link Bob to the criminal videos. \\
Noticeably, analyzing multimedia content shared on SMPs is not a trivial task. Indeed, besides stripping all metadata, SMPs usually re-encode images and videos. For example, Facebook policy is to down-scale and re-compress images so to obtain a target bit-per-pixel value \cite{MPBS15}; Youtube also scales and re-encodes digital videos \cite{giammarrusco2014source}. Needless to say, forensic traces left in the signal are severely hindered by such processing. Sensor pattern noise, however, is one of the most robust signal-level features, surviving down-scaling followed by compression. Nevertheless, when it comes to link the SPN extracted from, say, a Youtube video and a Facebook image, a new problem arises: since both content have been scaled/cropped by an unknown amount, such transformation must be estimated in order to align the patterns.
\begin{figure}
    \begin{center}
\includegraphics[width=\columnwidth]{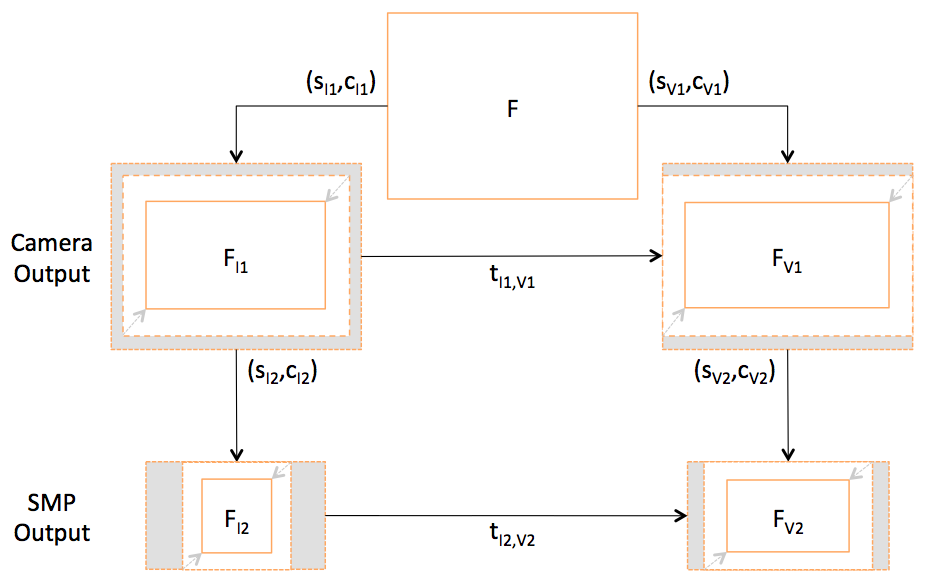}
    \caption{Geometric transformations applied to the sensor pattern from the full frame to the image and video outputs on both social media platforms.}
    \label{f:geomTransf}
    \end{center}
\end{figure}

Interestingly, the hybrid approach can be applied to this scenario. In Fig.~\ref{f:geomTransf} the geometric transformations occurring on the contents are summarized, starting from the full frame $F$; an image $F_{I_1}$ is produced by the acquisition process from $F$ with scaling and cropping factors $s_{I_1}$ and $c_{I_1}$ respectively. The uploading process over the SMP applies a new transformation - with factors $s_{I_2}$ and $c_{I_2}$ - thus producing $F_{I_2}$. In a similar way, the video $F_{V_1}$ is generated from the camera and $F_{V_2}$ is uploaded onto another SMP - with cropping and scaling factors of $s_{V_1}$, $c_{V_1}$ and $s_{V_2}$, $c_{V_2}$ respectively.
It can be easily deduced that, for both native and uploaded contents, image and video fingerprints are linked by a geometric transformation consisting in a cropping and scaling operation. Then, the hybrid approach that we used to determine the transformation $t_{I_1,V_1}$ which aligns the fingerprints of two native contents can be also applied to determine $t_{I_2,V_2}$, thus directly linking $F_{I_2}$ to $F_{V_2}$. \\
Two main drawbacks are expected for this second application. Firstly the compared contents have been probably compressed twice and the SPN traces are likely deteriorated. Furthermore it may be hard to guess the right scaling and cropping parameters just from $F_{I_2}$ and $F_{V_2}$. In these cases, an exhaustive search of all plausible scaling and cropping factors is required. In Section \ref{per:FaceTube} the proposed application is tested to link the images of a Facebook profile to the videos of a YouTube profile. \\



\section{Dataset for Hybrid Source Identification}
\label{sec:dataset}
We tested the proposed technique on an extensive dataset consisting of $1978$ flat field images, $3311$ images of natural scenes and $339$ videos captured by $18$ devices from different brands (Apple, Samsung, Huawei, Microsoft, Sony). YouTube versions of all videos and Facebook versions of all images (in both High and Low Quality) were also included. This dataset will be made available to the scientific community \footnote{\url{https://lesc.dinfo.unifi.it/en/datasets}}.
In the following we detail the dataset structure.
\subsection{Native contents}
We considered $18$ different modern devices, both smartphones and tablets.
Pictures and videos have been acquired with the default device settings that, for some models, include the automatic digital video stabilization. In Table \ref{tab:devices} we reported the considered models, their standard image and video resolution and whether the digital stabilization was active on the device.
From now on we'll refer to these devices with the names $C1, \dots, C18$ as defined in the Table~\ref{tab:devices}.
For each device we collected at least:
\begin{itemize}
 \item Reference side: $100$ flat-field images depicting skies or walls; $150$ images of indoor and outdoor scenes; $1$ video of the sky captured with slow camera movement, longer than $10$ seconds
 \item Query Side: videos of flat textures, indoor and outdoor scenes. For each of the video categories (flat, indoor and
outdoor) at least $3$ different videos have been captured
considering various scenarios: i) still camera, ii) walking
operator and iii) panning and rotating camera. We'll refer to them
as \emph{still}, \emph{move} and \emph{panrot} videos
respectively. Thus, each device has at least 9 videos, each one lasting more than $60$ seconds.
 \end{itemize}

\begin{table}[!ht]
    \centering
    \caption{Considered devices with their default resolution settings for image and video acquisition. 
    }
    \small
    \begin{tabular}{l @{\hspace{2mm}} l @{\hspace{2mm}} l @{\hspace{2mm}} l @{\hspace{2mm}} l  }
        \toprule
        ID & model & image          & video             & digital \\
             &              &   resolution  &   resolution  & stab \\
        \midrule
        C1  & Galaxy S3     & $3264 \times 2448$ & $1920 \times 1080$ & off \\
        C2  & Galaxy S3 Mini    & $2560 \times 1920$ & $1280 \times 720$  & off \\
        C3  & Galaxy S3 Mini    & $2560 \times 1920$ & $1280 \times 720$  & off \\
        C4  & Galaxy S4 Mini    & $3264 \times 1836$ & $1920 \times 1080$ & off \\
        C5  & Galaxy Tab 3 10.1 & $2048 \times 1536$ & $1280 \times 720$  & off \\
        C6  & Galaxy Tab A 10.1 & $2592 \times 1944$ & $1280 \times 720$  & off \\
        C7  & Galaxy Trend Plus & $2560 \times 1920$ & $1280 \times 720$  & off \\
        C8  & Ascend G6      & $3264 \times 2448$ & $1280 \times 720$  & off \\
        C9 & Ipad 2             & $960  \times 720$  & $1280 \times 720$  & off \\
        C10 & Ipad Mini         & $2592 \times 1936$ & $1920 \times 1080$ & on \\
        C11 & Iphone 4s         & $3264 \times 2448$ & $1920 \times 1080$ & on \\
        C12 & Iphone 5         & $3264 \times 2448$ & $1920 \times 1080$ & on \\
        C13 & Iphone 5c         & $3264 \times 2448$ & $1920 \times 1080$ & on \\
        C14 & Iphone 5c         & $3264 \times 2448$ & $1920 \times 1080$ & on \\
        C15 & Iphone 6          & $3264 \times 2448$ & $1920 \times 1080$ & on \\
        C16 & Iphone 6          & $3264 \times 2448$ & $1920 \times 1080$ & on \\
        C17 & Lumia 640          & $3264 \times 1840$ & $1920 \times 1080$ & off \\
        C18 &  Xperia Z1c          & $5248 \times 3936$ & $1920 \times 1080$ & on \\
        \bottomrule
    \end{tabular}
    \label{tab:devices}
\end{table}

\subsection{Facebook and YouTube sharing platforms}
\label{per:FaceTube}
Images have been uploaded on Facebook in both low quality (LQ) and high quality (HQ). The upload process eventually downscales the images depending on their resolutions and selected quality~\cite{MPBS15}.
Videos have been uploaded to YouTube through its web application and then downloaded through Clip Grab~\cite{clipgrab} selecting the best available resolution. 
\footnote{The metadata orientation has been removed from all of the images and videos to avoid unwanted rotation during the contents upload.}

\section{Experimental validation}
\label{sec:experiments}
The experimental section is split in four parts, each focused on a different contribution of the proposed technique:
\begin{enumerate}
    \item we determine the cropping and scaling parameters applied by each device model;
    \item we verify that, in the case of non-stabilized video, the performance of the hybrid approach is comparable with the source identification based on a video reference;
    \item we show the effectiveness in identifying the source of in-camera digitally stabilized videos;
    \item we show the performance in linking Facebook and YouTube profiles.
\end{enumerate}

\subsection{Fingerprints matching parameters}
\label{sec:fingmatch} The scaling and cropping factors applied by
each device were derived by registering the reference video
fingerprint $\widetilde{\mathbf{K}}_{V}$ on a reference
fingerprint $\widetilde{\mathbf{K}}_{I}$ derived from still images
according to the $P$ statistic (Eq.~\ref{eq:p}). For each device
we estimated $\widetilde{\mathbf{K}}_{I}$ by means of $100$ images
randomly chosen from the flat-field pictures. For non-stabilized
videos, $\widetilde{\mathbf{K}}_{V}$ was derived by means of the
first $100$ frames of the reference video available for that
device.
In Table~\ref{tab:registrations} we reported the obtained cropping
parameter (in terms of coordinates of the upper-left corner of the
cropped area along $x$ and $y$ axes, whereas the right down corner
is derived by the video size) and the scaling factor, maximising
the PCE. For instance, $C1$ image fingerprint should be scaled
with a factor $0.59$ and cropped on the upper left side of $307$
pixels along the $y$ axis to match the video fingerprint; $C9$ is
a pretty unique case in which the full frame is applied for video
and is left and right cropped of $160$ pixels to capture images.
\begin{table}[!ht]
    \centering
        \caption{Rescaling and cropping parameters that link image and video SPNs for the considered devices, in absence of in-camera digital stabilization.}
    \small
    \begin{tabular}{l @{\hspace{5mm}} r @{\hspace{5mm}} l @{\hspace{5mm}} l }
        \toprule
        ID      &   scaling &   central crop along  \\
                &            &  $x$ and $y$ axes   \\
        \midrule
        C1      & 0.59          & [0 307]  \\
        C2      & 0.5               & [0 228]   \\
        C3      & 0.5               & [0 228]   \\
        C4      & 0.59          & [0 0]     \\
        C5      & 1             & [408 354] \\
        C6      & 0.49      & [0 246]   \\ 
        C7      & 0.5       & [0 240]   \\ 
        C8      & 0.39          & [0 306]   \\  
        C9      & 1             & [-160 0]  \\
        C17    & 0.59       & [0 1]   \\
        \bottomrule
    \end{tabular}
    \label{tab:registrations}
\end{table}

In case of stabilized video, the cropping and scaling factors
varies in time with possible rotation applied too. For these
devices we thus determined the registration parameters of the
first $10$ frames of the available video reference; the main
statistics are reported in Table~\ref{tab:stabPar} .

\begin{table}[!ht]
    \centering
    \caption{Rescaling and cropping parameters that link image and video SPNs for the considered devices using in-camera digital stabilization. The values are computed on the
first $10$ frames of the available video reference; min, median (bold),
and max values are represented.}
    \begin{tabular}{l @{\hspace{1mm}} c @{\hspace{2mm}}  l @{\hspace{1mm}}  l @{\hspace{2mm}}  l }
        \toprule
        ID      &   scaling                 &   central crop along $x$ and $y$      & rotation (CCW)        \\
         \midrule
        C10      &     [0.806 \textbf{0.815} 0.821]   & [243 \textbf{256} 261] [86 \textbf{100} 103] & [-0.2 \textbf{0} 0.2] \\
        C11      &    [0.748 \textbf{0.750} 0.753]   &  [380 \textbf{388} 392] [250 \textbf{258} 265] & [-0.2 \textbf{0} 0.2]  \\
        C12      &     [0.684 \textbf{0.689} 0.691]  & [287 \textbf{294} 304] [135 \textbf{147} 165]  &  [-0.2 \textbf{0} 0.6] \\
        C13      &     [0.681 \textbf{0.686} 0.691]  & [301 \textbf{318} 327] [160 \textbf{181} 195] & [-0.4 \textbf{0} 1] \\
        C14      &     [0.686 \textbf{0.686} 0.689] &  [261 \textbf{301} 304] [119 \textbf{161} 165] & [-0.4 \textbf{0} 0] \\
        C15      &     [0.696 \textbf{0.703} 0.713] & [298 \textbf{322} 345] [172 \textbf{190} 218] & [-0.2 \textbf{0.2} 1.6]  \\
        C16      &      [0.703 \textbf{0.706} 0.708] & [315 \textbf{323} 333] [178 \textbf{187} 201] &  [-0.2 \textbf{0.2} 0.4] \\
        C18     &       [0.381 \textbf{0.384} 0.387] &   [548 \textbf{562} 574] [116 \textbf{121} 126]   & [0 \textbf{0} 0] \\
        \bottomrule
    \end{tabular}
        \label{tab:stabPar}
\end{table}

These data can be
exploited to reduce the parameter search space in case of
source identification of digitally stabilized videos. Indeed, an
exhaustive search of all possible scaling and rotation
parameters, required in a blind analysis, would be infeasible on
large scale: in our tests a totally blind search can take up to 10 minutes per frame on a standard computer, while the informed search reduces the time to less the a minute for stabilized videos and a few seconds for non-stabilized videos.

\subsection{HSI Performance}
\label{subs:res_native} 
\begin{figure}[h]
\begin{center}
\includegraphics[width=\columnwidth]{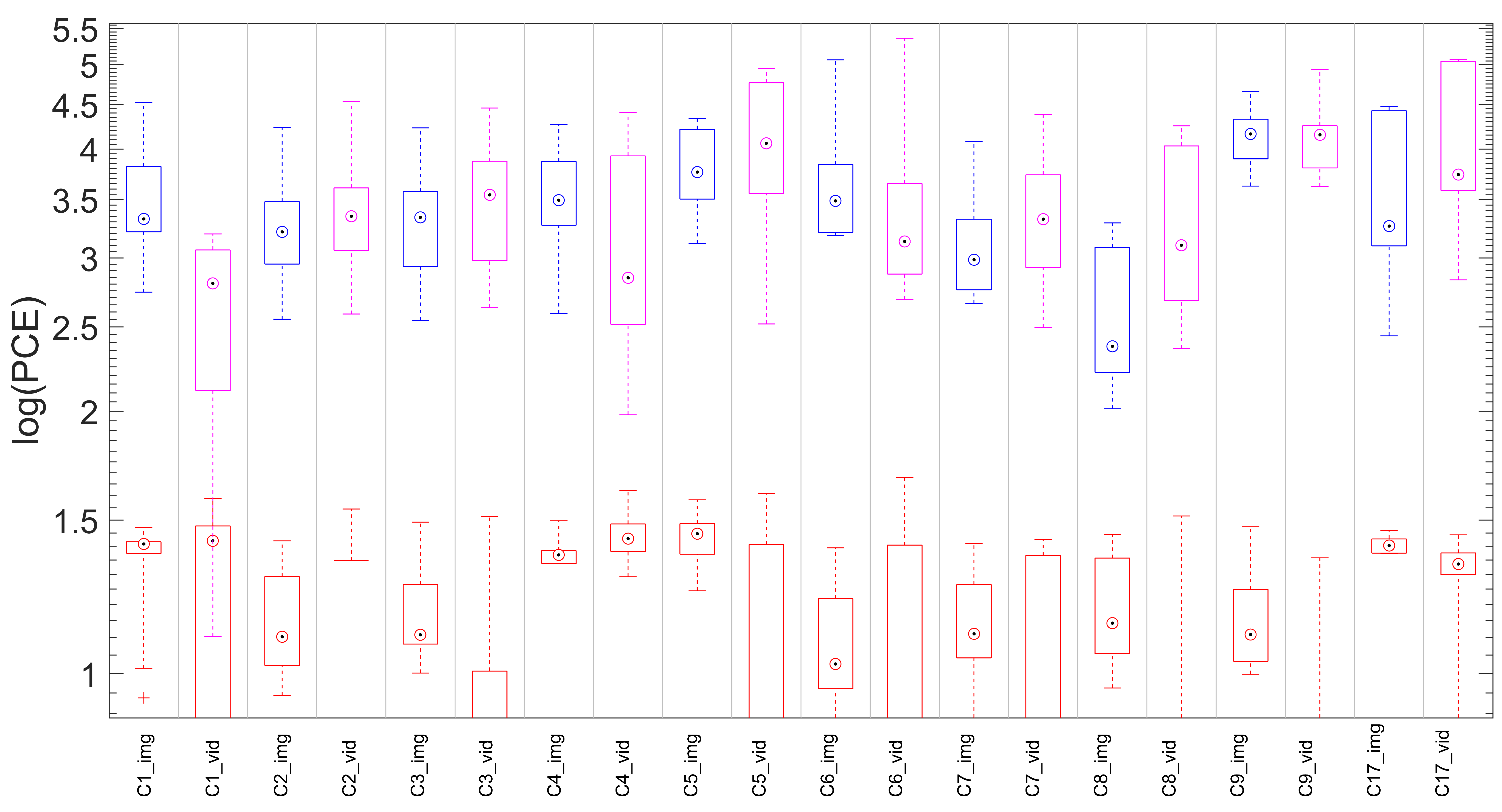}
\caption{(Best viewed in colors) Matching statistics $mP_I$ and
$mP_V$ are represented by the blue and pink boxplots,
respectively. Correspondent mismatching statistics in red. On each
box, the central mark indicates the median, and the bottom and top
edges of the box indicate the 25th and 75th percentiles,
respectively. The whiskers denote the minimum and maximum of the statistics. For plotting purposes, we defined $log(a) = -\infty \forall a\leq0$.
} \label{fig:HSItest}
\end{center}
\end{figure}

In this section we compare the proposed
technique with the state of the art approach, where the
fingerprint is derived estimating the SPN from a reference video. The comparison is only meaningful for non-stabilized devices. For each device, the reference
fingerprints $\widetilde{\mathbf{K}}_{I}$ and
$\widetilde{\mathbf{K}}_{V}$ were derived respectively from the
first $100$ natural reference images (for the proposed method) and from the
first $100$ of the
reference video (for the video reference approach). Given a video query, the fingerprint to be tested
was derived by the first $100$ frames and compared with
$\widetilde{\mathbf{K}}_{V}$ and with $\widetilde{\mathbf{K}}_{I}$
adopting the cropping and scaling parameters expected for the
candidate device (Eq.~\ref{eq:PCE}). We refer to the test
statistics as $P_V$ and $P_I$ to distinguish the reference origin
(video frames or still images). For each device we tested all
available matching pairs (reference and query from the same source
device) and an equal number of mismatching pairs (reference and
query from different source devices) randomly chosen from all
available devices. We refer to these statistics as $mP_I$ and
$mmP_I$ respectively ($mP_V$ and $mmP_V$ for video references). In
Fig.~\ref{fig:HSItest} we report for each device: i) the
statistics $mP_I$ and $mP_V$ (blue and pink respectively) of
matching pairs; ii)
$\overline{mmP}_I$ and $\overline{mmP}_V$ (in red), the statistics for mismatching cases. \\
The plot shows that distributions can be perfectly separated when the reference is estimated from images (100\% accuracy), while in the video reference case the accuracy is 99.5\%, confirming that the performance are comparable.

\subsection{HSI Performance on Stabilized Videos}
\label{subs:res_stab} State of the art results in identifying the source
of a stabilized video are provided in~\cite{2017stabilisedvideo}.
The authors, based on a similar registration protocol, analyze the
performance using both non-stabilized and stabilized references.
Their results are reported in Table~\ref{table:memon} for convenience: we see that, if a
non-stabilized reference is available, the method achieves a true
positive rate $0.83$. Unfortunately in several modern devices
(e.g., Apple smartphones) digital stabilization cannot be turned
off without third party applications; in this case, only
stabilized reference can be exploited, achieving a TPR of $0.65$.

\begin{figure}[h]
\begin{center}
\includegraphics[width=\columnwidth]{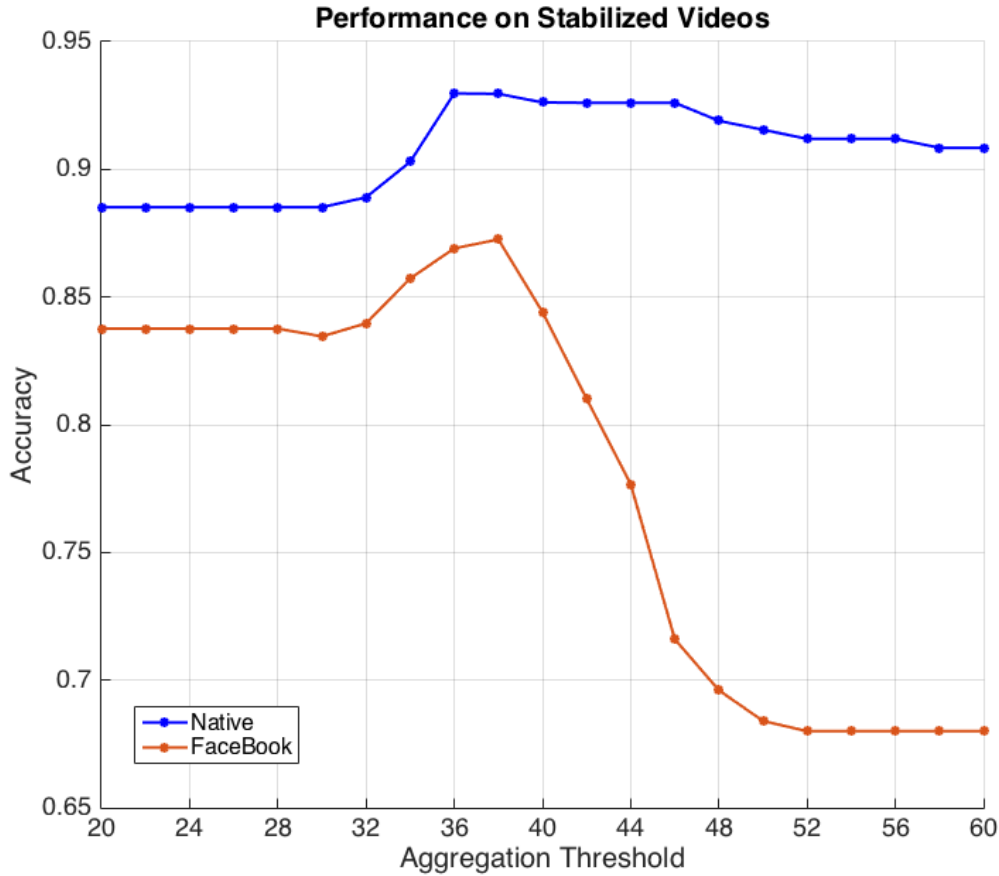}
\caption{(Best viewed in colors) Mean accuracy of source identification on digitally stabilized videos by varying the aggregation threshold $\tau$. Native and Facebook (HQ) contents are referred in blue and orange respectively.}
\label{fig:ResultsStabilized}
\end{center}
\end{figure}

\begin{figure}[!h]
\begin{center}
\includegraphics[width=\columnwidth]{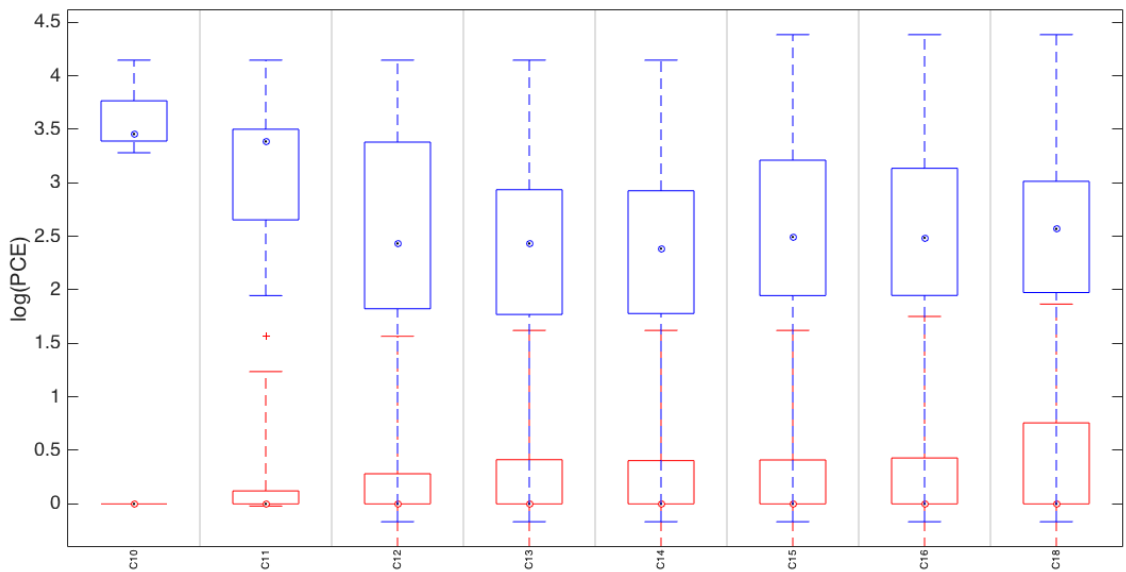}
\caption{(Best viewed in colors) Details of the performance
achieved with best aggregation threshold (38) on native stabilized
videos. Matching and mismatching statistics are reported in blue and red, respectively, for
each device.} 
\label{fig:HSItestStab}
\end{center}
\end{figure}

In the following, we will show that exploiting the proposed HSI
method, this performance drop can be solved.
\begin{table}[]
\centering
\caption{Performance of Source Identification of
digitally stabilized video (using ffmpeg) using both non-stabilized
and stabilized references reported in~\cite{2017stabilisedvideo}.}
\label{table:memon}
\begin{tabular}{  c  c  c  c }
\hline
  Reference & Query & TPR & FPR \\
  \hline
  Non-stabilized & Stabilized & 0.83 & 0 \\
  Stabilized & Stabilized & 0.65 & 0 \\
  \hline
\end{tabular}
\end{table}
For each device, the reference fingerprints
$\widetilde{\mathbf{K}}_{I}$ was estimated from $100$ natural
images. Given a video query, each frame is registered on
$\widetilde{\mathbf{K}}_{I}$ searching within the expected
parameters for the candidate device (as derived in
Section~\ref{sec:fingmatch}). The video fingerprint
$\widetilde{\mathbf{K}}_{V}$ is then obtained by aggregating all registered video
frames whose PCE wrt $\widetilde{\mathbf{K}}_{I}$ overcomes the aggregation threshold $\tau$. Finally, the aggregated fingerprint $\widetilde{\mathbf{K}}_{V}$ is compared with the reference SPN $\widetilde{\mathbf{K}}_{I}$. All
tests were performed limiting the analysis to the first $5$ frames of each video.
For each device, we tested all available matching videos and an equal number of mismatching videos randomly chosen from all available devices. In Figure~\ref{fig:ResultsStabilized} we show the system accuracy by varying the aggregation threshold $\tau$.
Table~\ref{table:HSIstab} shows, for different values of $\tau$, the the TPR and FPR corresponding to the best accuracy. Fig.~\ref{fig:HSItestStab} shows the matching and mismatching PCE statistics obtained using $\tau=38$.\\
Results clearly show that, using $\tau = 50$, the system achieves a TPR equal to
$0.83$, which is totally consistent with results achieved
in~\cite{2017stabilisedvideo}, but in our case without the need for a non-stabilized video
reference. Moreover, results show that using a slightly lower aggregation threshold some
improvement can be achieved, (TPR $0.86$ with an aggregation
threshold of $38$).

\begin{table}[]
\centering \caption{Performance of the proposed method for different values of the aggregation threshold $\tau$.}
\begin{tabular}{ccccc}
\hline
Aggregation & Accuracy & TPR  & FPR   & AUC  \\
threshold ($\tau$) &      &          &       &  \\
\hline
30                &89\%    & 0.79 & 0.02     & 0.93 \\
32                &89\%    & 0.82 & 0.05      & 0.94 \\
34                &90\%    & 0.84 & 0.03       & 0.94 \\
36                &93\%    & 0.87 & 0.01      & 0.95 \\
38                &93\%    & 0.86 & 0         & 0.94 \\
40                &93\%    & 0.87 & 0.01      & 0.93 \\
42                &93\%    & 0.85 & 0         & 0.93 \\
44                &93\%    & 0.85 & 0         & 0.93 \\
46                &93\%    & 0.85 & 0         & 0.93 \\
48                &92\%    & 0.84 & 0         & 0.92 \\
50                &92\%    & 0.83 & 0        & 0.92 \\
52                &91\%    & 0.82 & 0        & 0.91 \\
54                &91\%    & 0.82 & 0         & 0.91 \\
\hline
\end{tabular} \label{table:HSIstab}
\end{table}

\subsection{Results on contents from SMPs}
\begin{figure}[h!b]
\begin{center}
\includegraphics[width=0.7\columnwidth]{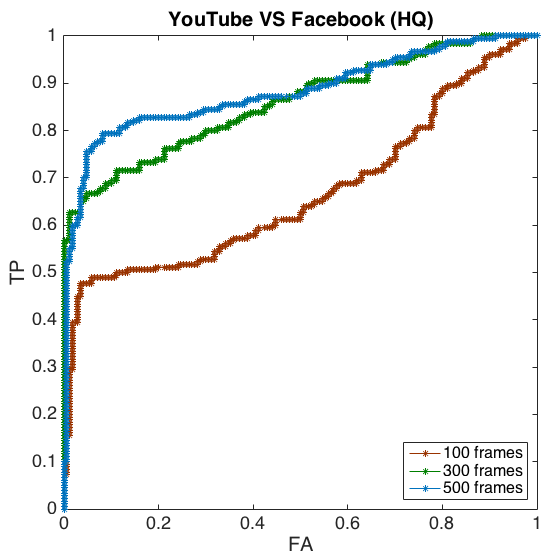}
\caption{(Best viewed in colors) ROC curve for profile linking
between non-stabilized YouTube videos and Facebook HQ images by
varying the number of frames to estimate the video reference.}
\label{fig:ROChq}
\end{center}
\end{figure}

In this section we test the HSI approach in the application scenario of linking Facebook and YouTube accounts containing images and videos captured with the same device. For clarity we considered
the non-stabilized and stabilized cases separately. Furthermore,
we conducted two experiments: one estimating the SPN from images uploaded to Facebook using the high-quality option, and another experiment estimating the SPN from images uploaded using the low quality  option. A detailed explanation of the differences between the two options is given in ~\cite{MPBS15}, here we only mention the fact that under low-quality upload images are downscaled so that their maximum dimension does not exceed 960 pixels, while under high-quality upload the maximum allowed dimension rises to 2048 pixels.
Throughout all tests, we used $100$ images for estimating the camera fingerprint. After estimating
image and video fingerprints according to the method described in previous sections,
we investigated the matching performance by varying the number of frames employed to estimate the  fingerprint of the query video. For sake of simplicity we reported the
aggregated results with a ROC curve where true positive rate
and false alarm rate are compared, and we used the AUC as an overall index of
performance. Similarly to the previous experiment, we considered all
available matching videos for each device (minimum 9 videos, 17 on average) and an equal number of
randomly selected mismatching videos. In Fig.~\ref{fig:ROChq} we
report the results of the first experiment (high quality Facebook reference
vs YouTube non-stabilized videos) by using $100$, $300$ and $500$
frames to estimate the fingerprint from the video. It can be easily
noticed that a hundred frames is rarely enough to correctly link
two profiles. Moving from $100$ to $300$ frames significantly
improves the performance, and much slighter improvement can be
achieved passing from $300$ to $500$ frames. The AUC values for
the three cases are $0.67$, $0.86$, $0.88$, respectively.
\begin{figure}[h!]
\begin{center}
\includegraphics[width=0.7\columnwidth]{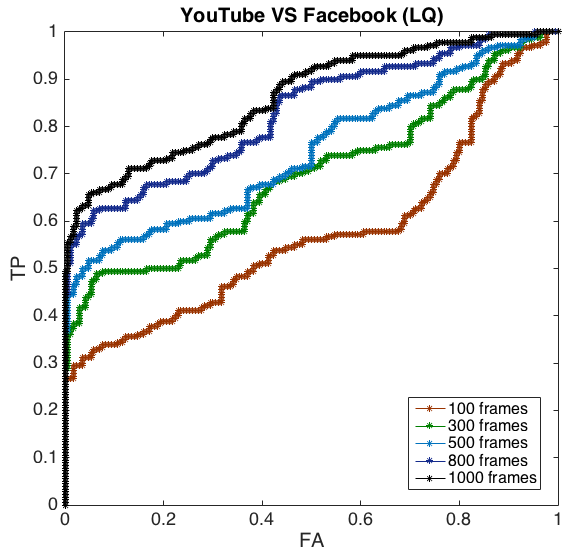}
\caption{(Best viewed in colors) ROC curve for profile linking
between non-stabilized  YouTube videos and Facebook LQ images by
varying the number of frames to estimate the video reference.}
\label{fig:ROClq}
\end{center}
\end{figure}

When Facebook images uploaded in low quality are used as reference, the estimated pattern is expected to be less reliable than for the high quality case.  This degradation on the reference side can be mitigated by using more robust estimates on the query side; for this reason, for the low quality case, we also considered using $500$, $800$ and
$1000$ frames for extracting the query pattern, achieving ROC curves reported in
Fig.~\ref{fig:ROClq}. The corresponding AUC values are $0.57,
0.70, 0.75, 0.83$ and $0.86$ using $100, 300, 500, 800$ and $1000$
frames respectively.

\begin{figure}[h!]
\begin{center}
\includegraphics[width=\columnwidth]{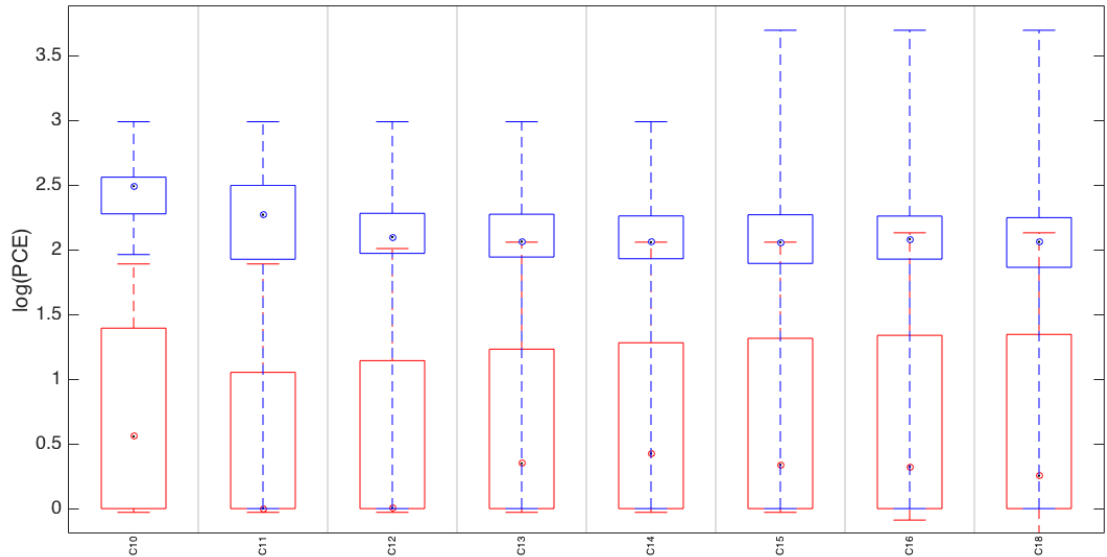}
\caption{(Best viewed in colors) Details of the performance achieved with best aggregation threshold (38) on stabilized YouTube videos using Facebook (HQ) images as references. Matching and mismatching statistics are reported in blue and red, respectively, for each of the devices.}
\label{fig:HSItestFBYT}
\end{center}
\end{figure}

Let us now focus on the case of in-camera stabilized  videos downloaded from Youtube. Fig.~\ref{fig:ResultsStabilized} reports the achieved performance for different values of the aggregation threshold $\tau$. The plot suggests that using $\tau = 38$ remains the best choice also in this experiment, leading to 87.3\% overall accuracy. Fig.\ref{fig:HSItestFBYT}  details the performance for each device by applying such aggregation threshold. Thus, we can say that the hybrid approach to source identification provides promising results for linking SMP profiles even in the case of in-camera digitally stabilized videos.

\section{Conclusions}
\label{sec:conclusion} In this paper we proposed an hybrid approach to video source identification using a reference fingerprint derived from still images. We showed that the hybrid approach yields comparable or even better performance than the current strategy of using a video reference in the case of non-stabilized videos. As a major contribution, our approach allows reliable source identification even for videos produced by devices that enforce digital in-camera stabilization (e.g., all recent Apple devices), for which a non-stabilized reference is not available. 
We reported the geometrical relationships between image and video acquisition process of $18$ different devices, even in case of digitally stabilized videos.
The proposed method was applied to
link image and video contents belonging to different social media
platforms: its effectiveness has been
proved to link Facebook images to YouTube videos, with promising results even in the case of digitally stabilized videos. Specifically,
when low quality Facebook images are involved, we showed that some hundreds
of video frames are required to effectively link the two sensor
pattern noises. We performed experiments on an brand
new dataset of $339$ videos and $5289$ images from $18$ different
modern smartphones and tablets, each accompanied by its Facebook
and YouTube version. The dataset will be shared with the research
community to support advancements on these topics.
The main limitation of the proposed approach is the need for a brute force search for determining scale (and, in the case of stabilized devices, rotation) when no information on the tested device is available. A possible way to mitigate this problem would be to design SPN descriptors that are simultaneously invariant to crop and scaling. This challenging task is left for future work.
\bibliographystyle{IEEEtran}
\bibliography{mybibfile}


\end{document}